\documentclass[sigconf, nonacm]{acmart}
\renewcommand\footnotetextcopyrightpermission[1]{}

\AtBeginDocument{%
  }

\setcopyright{acmlicensed}
\copyrightyear{2018}
\acmYear{2018}
\acmDOI{XXXXXXX.XXXXXXX}
\acmConference[Conference acronym 'XX]{Make sure to enter the correct
  conference title from your rights confirmation email}{June 03--05,
  2018}{Woodstock, NY}
\acmISBN{978-1-4503-XXXX-X/2018/06}

\usepackage{enumitem}
\usepackage{xcolor}
\usepackage{diagbox}
\usepackage{amsmath}
\usepackage{multirow}

\usepackage{algorithmic}
\usepackage[linesnumbered,ruled,vlined]{algorithm2e}

\usepackage{booktabs}
\usepackage{array}
\usepackage{tabularx}

\usepackage{xspace}
\usepackage{pifont}
\newcommand{\cmark}{\ding{51}}
\newcommand{\xmark}{\ding{55}}

\usepackage[table]{xcolor}

\usepackage{xspace}
\newcommand{\AVD}{\texttt{AVD}\xspace}

\usepackage{graphicx}    
\usepackage{subcaption}

\usepackage{geometry}   
\usepackage{float}      
\usepackage{caption}

\begin{document}

\title{Adaptive Value Decomposition: Coordinating a Varying Number of Agents in Urban Systems}

\author{Yexin Li}
\affiliation{%
  \institution{State Key Laboratory of General Artificial Intelligence, BIGAI}
  \city{Beijing}
  \country{China}
}
\email{liyexin@bigai.ai}

\author{Jinjin Guo}
\affiliation{%
  \institution{JD.com}
  \city{Beijing}
  \country{China}}
\email{umguojinjin@outlook.com}

\author{Haoyu Zhang}
\affiliation{%
  \institution{Beijing Technology and Business University}
  \city{Beijing}
  \country{China}
}

\author{Yuhan Zhao}
\affiliation{%
 \institution{State Key Laboratory of General Artificial Intelligence, BIGAI}
 \city{Beijing}
 \country{China}}

\author{Yiwen Sun}
\affiliation{%
  \institution{Peking University}
  \city{Beijing}
  \country{China}}

\author{Zihao Jiao}
\affiliation{%
  \institution{Beijing Technology and Business University}
  \city{Beijing}
  \country{China}}

\renewcommand{\shortauthors}{Yexin Li et al.}

\begin{abstract}
  Multi-agent reinforcement learning (MARL) provides a promising paradigm for coordinating multi-agent systems (MAS). However, most existing methods rely on restrictive assumptions, such as a fixed number of agents and fully synchronous action execution. These assumptions are often violated in urban systems, where the number of active agents varies over time, and actions may have heterogeneous durations, resulting in a semi-MARL setting. Moreover, while sharing policy parameters among agents is commonly adopted to improve learning efficiency, it can lead to highly homogeneous actions when a subset of agents make decisions concurrently under similar observations, potentially degrading coordination quality. To address these challenges, we propose \textbf{Adaptive Value Decomposition} (\AVD), a cooperative MARL framework that adapts to a dynamically changing agent population. \AVD further incorporates a lightweight mechanism to mitigate action homogenization induced by shared policies, thereby encouraging behavioral diversity and maintaining effective cooperation among agents. In addition, we design a training–execution strategy tailored to the semi-MARL setting that accommodates asynchronous decision-making when some agents act at different times. Experiments on real-world bike-sharing redistribution tasks in two major cities, London and Washington, D.C., demonstrate that \AVD outperforms state-of-the-art baselines, confirming its effectiveness and generalizability.
\end{abstract}

\ccsdesc[500]{Information systems~Spatial-temporal systems}

\keywords{Multi-Agent Systems, Multi-Agent Reinforcement Learning, Dynamic Agent Population, Value Decomposition, Action Homogenization, Asynchronous Decision-Making}



\maketitle

\section{Introduction}

Urban multi-agent systems (MAS) are increasingly deployed to support public services such as transportation, logistics, and shared mobility. A representative instance is the fleet of redistribution vehicles in bike-sharing operations~\cite{liu2016bike, li2018bike}, where the inherently spatio-temporal imbalance of bike demand~\cite{li2015bike, li2020bike} necessitates dynamic bike redistribution. In this context, redistribution vehicles can naturally be modeled as agents. However, designing long-term decision-making strategies for these agents remains challenging.

Reinforcement learning (RL) has been widely applied to sequential decision-making problems in complex systems~\cite{StarCraft, pan2019bike, li2019logistics}. However, modeling a MAS using a centralized RL framework, where a single controller governs all agents, leads to an exponentially growing joint action space, rendering the problem computationally intractable. Cooperative multi-agent reinforcement learning (MARL) under the centralized training and decentralized execution (CTDE) paradigm~\cite{foerster2018counterfactual, rashid2020monotonic, yu2022surprising} provides a natural alternative. Under this paradigm, agents are trained in a centralized manner but execute their policies independently, enabling decentralized decision-making while preserving coordinated behavior.

Despite these advantages, applying MARL to urban MAS poses several practical challenges.

\begin{itemize}[leftmargin=1.5em]
    \item \textbf{Dynamic agent population}. Due to changing operational demands, the number of active agents can vary over time—more agents may be required during peak periods than during off-peak periods, while unexpected events or incidents can further increase or decrease the active workforce. Such variability poses challenges for representative MARL algorithms~\cite{rashid2020monotonic, foerster2018counterfactual}, which typically assume a fixed number of agents. A straightforward approach is to assume a maximum number of agents and handle cases with fewer active agents through padding. However, this strategy leads to inefficient use of computational resources and may still fail when the actual number of agents exceeds the predefined maximum.

    \item \textbf{Variable action durations}. In urban systems, agents often execute actions with variable durations~\cite{ruan2026agricultural}, leading to asynchronous behaviors and giving rise to a semi-MARL setting.

    \item \textbf{Homogeneous actions}. MARL methods often adopt parameter sharing across agents~\cite{chen2020light, li2020logistics}. Although efficient, shared policies can produce highly homogeneous actions when a subset of agents make decisions concurrently under similar observations. This issue is particularly pronounced in urban systems, where agents are often guided by demand signals that are concentrated in high-activity areas. As a result, these agents are more likely to encounter highly similar conditions at the same decision time, which further amplifies behavioral homogenization and degrades coordination quality in practice.
\end{itemize}

\textbf{Adaptive Value Decomposition} (\AVD) is a cooperative MARL framework designed to coordinate a variable number of agents. Building on the value decomposition technique~\cite{sunehag2017value, rashid2020monotonic}, \AVD employs agent-wise networks together with a hypernetwork that dynamically adapts to the current active agent set. To mitigate action homogenization caused by shared agent-wise policies, agents are initialized with individualized states, and lightweight stochastic perturbations are continuously injected into their real-time states at each decision step, encouraging behavioral diversity. For the semi-MARL setting with asynchronous action execution, we develop a CTDE-based training–execution strategy that enables agents to make independent decisions asynchronously while maintaining global coordination.

\AVD is evaluated on real-world bike-sharing datasets from two major cities, London and Washington, D.C. Extensive experimental results indicate that \AVD significantly outperforms representative baselines, demonstrating its robustness and effectiveness in realistic urban MAS. Moreover, zero-shot evaluation shows that policies learned by \AVD transfer effectively to scenarios with previously unseen agent populations, emphasizing its adaptability. Ablation studies further confirm the necessity of the lightweight mechanism in mitigating action homogenization.

In summary, our main contributions are as follows:

\begin{itemize}[leftmargin=1.5em]
    \item \textbf{Algorithmic innovation}. We propose \textbf{Adaptive Value Decomposition} (\AVD), a cooperative MARL framework that adapts to a variable number of active agents and incorporates a lightweight mechanism to mitigate action homogenization induced by shared policies.

    \item \textbf{Semi-MARL training–execution strategy}. A CTDE-based training–execution strategy is designed to accommodate asynchronous action execution, enabling agents to make independent decisions while maintaining coordinated behavior.

    \item \textbf{Empirical validation}. Extensive experiments on bike-sharing datasets from two cities show that \AVD significantly outperforms representative baselines, demonstrating its robustness and effectiveness in realistic urban MAS.

    \item \textbf{Benchmark contribution}. We develop and release a high-fidelity bike-sharing simulator to serve as a standardized benchmark for studying resource redistribution and multi-agent coordination in urban MAS.
\end{itemize}

\section{Related Work}
In this section, we review the literature on reinforcement learning, covering both single-agent and multi-agent methods, as well as decision-making in urban systems.

\subsection{Reinforcement Learning}

RL algorithms, including single-agent methods such as DQN~\cite{DQN}, PPO~\cite{PPO}, SAC~\cite{SAC}, DDPG~\cite{DDPG}, TD3~\cite{TD3}, IMPALA~\cite{IMPALA}, DSAC~\cite{DSAC, DSACT}, and CAE~\cite{li2025cae}, as well as multi-agent methods such as VDN~\cite{sunehag2017value}, QMIX~\cite{rashid2020monotonic}, COMA~\cite{foerster2018counterfactual}, MAPPO~\cite{yu2022surprising}, and ADAPT~\cite{zhang2026adapt}, have demonstrated strong performance in diverse domains, including Atari games~\cite{atari_dqn, DQN}, Go~\cite{Go}, StarCraft~\cite{StarCraft}, Civilization~\cite{qi2024civrealm}, and indoor navigation~\cite{szot2021habitat}.

Practical applications of RL often face challenges such as large action spaces~\cite{Gabriel2016large} and scalability limitations. In this work, a cooperative MARL framework is adopted to generate decentralized actions for individual agents, alleviating the exponential growth of the joint action space. Despite this, conventional MARL algorithms~\cite{rashid2020monotonic, foerster2018counterfactual} still struggle with several practical challenges, including the time-varying number of active agents, heterogeneous action durations, and the risk of action homogenization under shared agent-wise policies. These challenges form the focus of the present work.

\subsection{Decision-making in Urban Systems}

Decision-making problems in urban environments can generally be categorized into static and dynamic settings. In static decision-making, decisions remain fixed over extended periods. Representative examples include facility location planning to maximize trajectory coverage~\cite{traj2018}, and delivery zone partitioning to balance courier workloads based on historical demand patterns~\cite{Guo2023equitable}.

In contrast, dynamic decision-making requires actions to adapt to evolving system states on a continuous basis. RL has therefore been increasingly applied to a wide range of dynamic urban decision-making problems, including logistics and fleet management~\cite{li2019logistics, li2020logistics}, order dispatching~\cite{sadeghi2022reinforcement, sun2022optimizing, yue2024end}, urban sensing~\cite{wang2024urban}, traffic signal control~\cite{chen2020light}, and bike redistribution, among others.

Among these applications, bike-sharing operations~\cite{li2015bike, li2020bike} have attracted substantial research attention and also serve as the experimental domain in this work. Existing bike redistribution strategies can be broadly categorized into static redistribution~\cite{liu2016bike}, dynamic redistribution~\cite{li2018bike, yin2023bike, jing2024bike, liang2024bike}, and user-based repositioning~\cite{pan2019bike}. Static redistribution typically operates during non-operational periods. User-based repositioning incentivizes riders to rent from or return bikes to designated stations. In contrast, dynamic redistribution continuously adjusts redistribution decisions in response to real-time demand fluctuations and has become increasingly prevalent in both research and practice.

\section{Problem Statement}

Urban platforms can continuously monitor the real-time status of agents through low-overhead communication and tracking mechanisms commonly available in urban services—such as GPS-based location tracking and mobile network communication used in bike redistribution, ride-sharing, and logistics order dispatching. Given this capability, we adopt a fully observable MARL setting in this work, where each agent has access to the global state. Nonetheless, the proposed \AVD framework can be readily extended to partially observable MARL settings.

\paragraph{Semi-MARL Formulation} 
A cooperative multi-agent task in an urban system is defined by the tuple $(S, A, \mathcal{V}, \mathbb{P}, r, \gamma)$, where $S$ denotes the global state space, $A$ is the action space for each agent, and $\mathcal{V}$ represents the set of all potential agents, whose cardinality is unbounded. At each time step $\tau$, the set of active agents is denoted by $\mathcal{V}_{\tau} \subseteq \mathcal{V}$. Within this set, $V_{\tau} \subseteq \mathcal{V}_{\tau}$ comprises agents that have completed their previous actions, while the remaining agents, $\mathcal{V}{\tau} \setminus V_{\tau}$, have not yet completed them.

At time step $\tau$, the global state is denoted as $s_{\tau}$. Each active agent $v \in \mathcal{V}_{\tau}$ selects an action according to $a_{\tau}^{v} \sim \pi_{v}(\cdot \mid s_{\tau})$, where $\pi_{v}$ represents its policy. Upon executing the joint action $\boldsymbol{a}_{\tau} = (a_{\tau}^{v})_{v \in \mathcal{V}_\tau}$, the environment transitions to the next state according to $s_{\tau+1} \sim \mathbb{P}(\cdot \mid s_{\tau}, \boldsymbol{a}_{\tau})$ and provides an immediate reward $r_{\tau} = r(s_{\tau}, \boldsymbol{a}_{\tau})$ shared among all active agents in $\mathcal{V}_\tau$. The objective is to learn a set of policies $\{\pi_1, \pi_2, \ldots, \pi_{|\mathcal{V}|}\}$ that maximize the expected cumulative return over an episode of horizon $H$, defined as
\begin{equation}
    R = \sum_{\tau=1}^{H} \gamma^{\tau-1} r_{\tau}
    \label{eq:return}
\end{equation}
where $\gamma \in [0,1]$ is the discount factor.

\paragraph{Concrete Example of State, Action, and Reward}
To illustrate the definitions of state, action, and immediate reward in a concrete setting, we consider the bike redistribution task and describe each component in detail below.

\textbf{State}. At each time step $\tau$, the state is represented as $s_{\tau} = (s_{\tau}^{\text{env}}, s_{\tau}^{\text{agent}})$, comprising two components:

\begin{itemize}[leftmargin=1.5em]
    \item \textbf{Environment state} $s_{\tau}^{\text{env}} \in \mathbb{R}^{1 \times d_1}$ encodes the system-level information, including the inventory levels at all stations and the current time step.
    \item \textbf{Agent state} $s_{\tau}^{\text{agent}} \in \mathbb{R}^{|\mathcal{V}_\tau| \times d_2}$ represents the status of each active vehicle $v \in \mathcal{V}_\tau$. It includes the vehicle's current observation, which consists of its current or target station and its load, as well as the progress of its ongoing action. Action progress is expressed as a normalized fraction in $[0,1]$. For example, if an action requires $\Delta$ steps, the first step corresponds to $\frac{1}{\Delta}$.
\end{itemize}

\textbf{Action}. At time step $\tau$, the action of vehicle $v \in \mathcal{V}_\tau$ is $a_{\tau}^{v} = \{ g_{\tau}^{v}, m_{\tau}^{v} \}$, where $g_{\tau}^{v}$ is the target station, and $m_{\tau}^{v} \in \mathbb{Z}$ indicates the number of bikes to load or unload. Specifically, if $m_{\tau}^{v} \geq 0$, it corresponds to loading bikes, whereas if $m_{\tau}^{v} < 0$, it corresponds to unloading bikes. Actions may span multiple time steps depending on the distance to the target station and the workload.

\textbf{Reward}. At each time step $\tau$, the immediate reward $r_{\tau}$ is defined as the total number of bikes successfully rented across all stations during that step. It is shared among all active vehicles $\mathcal{V}_\tau$ at time $\tau$, reflecting the cooperative nature of the task.

\begin{figure*}[t]
    \centering
    \includegraphics[width=\linewidth, trim=0cm 9.0cm 0cm 10.0cm, clip]{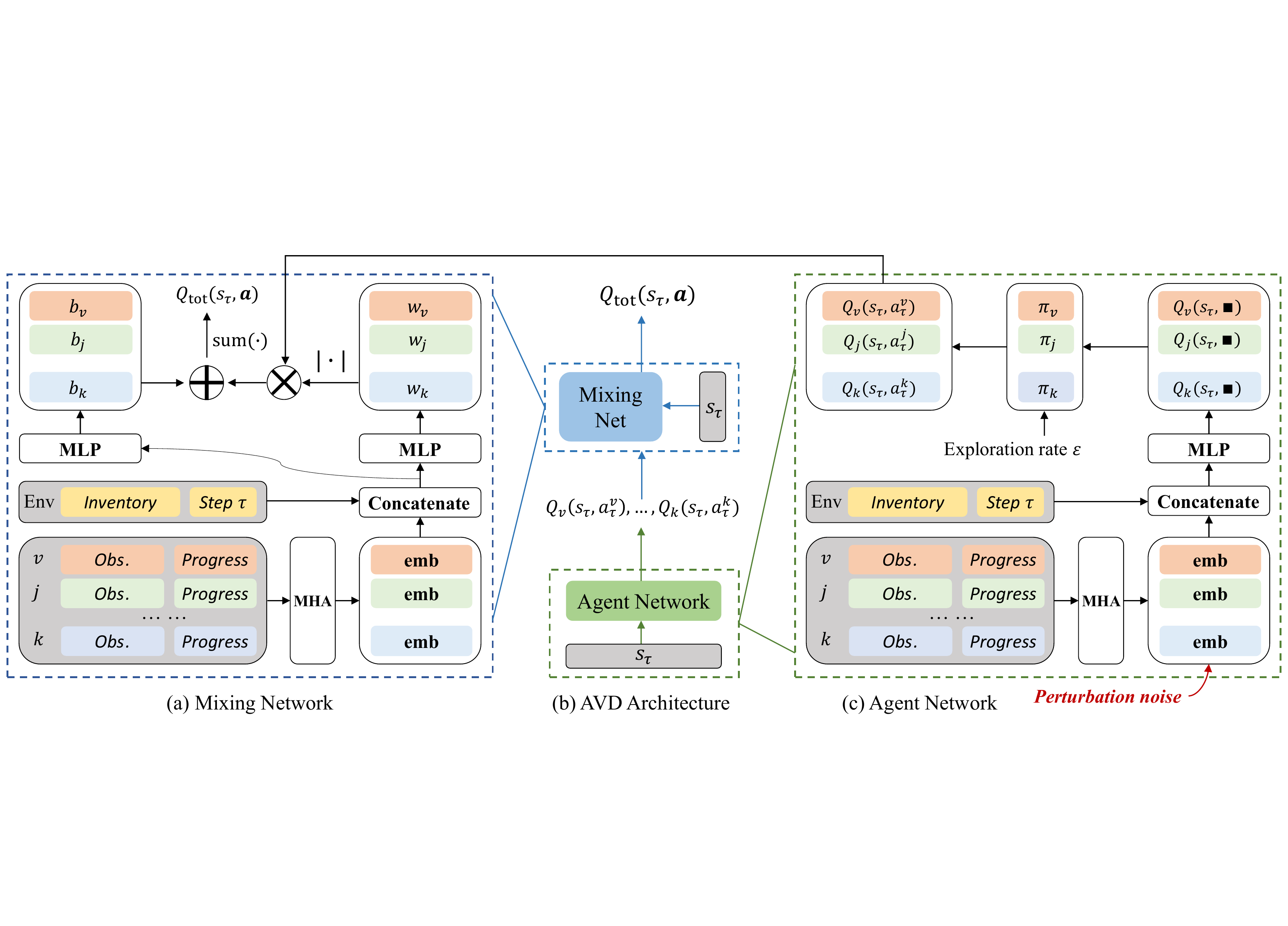}
    \caption{Illustration of \AVD, where $v, j, k \in \mathcal{V}_\tau$ denote active agents. Panel (b) illustrates the overall architecture of AVD. The current state $s_{\tau}$ is input to the shared agent network, as depicted in panel (c), which outputs the action-values $Q_{v}(s_{\tau}, a_{\tau}^{v}), Q_{j}(s_{\tau}, a_{\tau}^{j}), ..., Q_{k}(s_{\tau}, a_{\tau}^{k})$ for each agent under their selected actions $a_{\tau}^{v}, a_{\tau}^{j}, ..., a_{\tau}^{k}$. These action-values, along with the state $s_{\tau}$, are then fed into the mixing network, as shown in panel (a). The mixing network generates the weights $w_{v}, w_{j}, ..., w_{k}$ and biases $b_{v}, b_{j}, ..., b_{k}$ for the agents, and computes the total action-value using Eq.~\ref{eq:total_value}.}
    \label{fig:AVD}
\end{figure*}

\section{Methodology}
In light of the challenges inherent in urban MAS, including a time-varying agent population, action homogenization induced by shared policies, and heterogeneous action durations, we introduce \AVD to tackle the first two issues, and design a training–execution strategy to accommodate the third.

\subsection{Adaptive Value Decomposition}

In this subsection, motivated by QMIX~\cite{rashid2020monotonic}, we propose \AVD to learn the optimal policies $\pi_{1}, \pi_{2}, \dots, \pi_{|\mathcal{V}|}$ for the set of potential agents $\mathcal{V}$. Notably, the number of active agents $|\mathcal{V}_{\tau}|$ at each time step $\tau$ varies, whereas most existing MARL algorithms~\cite{foerster2018counterfactual, rashid2020monotonic} assume a fixed agent set.


Fig.~\ref{fig:AVD} illustrates the framework of \AVD. For an active agent $v \in \mathcal{V}_{\tau}$ at time step $\tau$, the agent network approximates its action-value function $Q_{v}(s_{\tau}, a_{\tau}^{v})$, which estimates the expected cumulative return for agent $v$ from time step $\tau$ to the end of the episode. The mixing network~\cite{rashid2020monotonic} aggregates the individual agent value functions $\left\{ Q_{v}|v \in \mathcal{V}_{\tau} \right\}$ using a state-dependent monotonic mixing function, as shown in Eq.~\ref{eq:mix_net}, to estimate the global action-value function representing the team’s overall return.

\begin{equation}
    Q_{\textup{tot}} (s_{\tau}, \boldsymbol{a}_{\tau}) = f \Big( \{ Q_v(s_{\tau}, a_{\tau}^{v}) \}_{v \in \mathcal{V}_\tau}; s_{\tau} \Big)
    \label{eq:mix_net}
\end{equation}

Notably, the mixing function satisfies the monotonicity constraint, i.e., $\frac{\partial f}{\partial Q_{v}} \ge 0$ for $\forall v \in \mathcal{V}_{\tau}$. This ensures that increasing an individual agent’s value does not decrease the total value, thereby enabling decentralized execution after centralized training. As illustrated in Eq.~\ref{eq:qmix}, the optimal joint action corresponds to the combination of each agent’s individually optimal action.

\begin{equation}
\operatorname*{argmax}_{\boldsymbol{a}_{\tau}} Q_{\textup{tot}} (s_{\tau}, \boldsymbol{a}_{\tau}) =
\begin{pmatrix}
\displaystyle \operatorname*{argmax}_{a_{\tau}^{v}} Q_{v}(s_{\tau}, a_{\tau}^{v})
\end{pmatrix}_{v \in \mathcal{V}_\tau}
\in \mathbb{R}^{|\mathcal{V}_\tau| \times 1}
\label{eq:qmix}
\end{equation}

\subsubsection{Architecture of the agent network}

Process agent state $s_{\tau}^{\text{agent}}$ using multi-head attention (MHA) layers to generate contextualized embeddings as shown in Eq.~\ref{eq:state_agent}. By leveraging MHA, each agent's embedding captures semantic relationships with other agents.

\begin{equation}
    \mathcal{H}_{\text{agent}}^{V} = \text{MHA}_{\text{agent}} (s_{\tau}^{\text{agent}}) \in \mathbb{R}^{|\mathcal{V}_{\tau}| \times d}
    \label{eq:state_agent}
\end{equation}

Broadcast system-level information $s_{\tau}^{\text{env}}$ to active agents to form the enriched representation $\mathcal{E}_{\text{agent}}$
\begin{align}
    & \mathcal{H}^{\text{env}} = \boldsymbol{1}_{|\mathcal{V}_{\tau}|} \otimes s_{\tau}^{\text{env}} \in \mathbb{R}^{|\mathcal{V}_{\tau}| \times d_{1}} \\
    & \mathcal{E}_{\text{agent}} = \text{concat}\left( \mathcal{H}^{\text{env}}, \mathcal{H}_{\text{agent}}^{V} \right) \in \mathbb{R}^{|\mathcal{V}_{\tau}| \times (d_{1} + d)}
\end{align}
where $\boldsymbol{1}_{|\mathcal{V}_{\tau}|}$ is a length-$|\mathcal{V}_{\tau}|$ vector of ones, and $\otimes$ denotes the Kronecker product.

A multi-layer perceptron (MLP) is then applied to estimate the action values for each agent over all possible actions, as shown in Eq.~\ref{eq:individual_value}, where the MLP is applied row-wise to $\mathcal{E}_{\text{agent}}$.

\begin{equation}
    \Big( Q_v \Big)_{v \in \mathcal{V}_\tau} = \text{MLP}_{\text{agent}} (\mathcal{E}_{\text{agent}}) \in \mathbb{R}^{|\mathcal{V}_{\tau}| \times |A|}
    \label{eq:individual_value}
\end{equation}

\subsubsection{Architecture of the mixing network}

Following a similar design as the agent network, obtain agent state embeddings using an MHA, as defined in Eq.~\ref{eq:state_mix}.

\begin{equation}
    \mathcal{H}_{\text{mixing}}^{V} = \text{MHA}_{\text{mixing}} (s_{\tau}^{\text{agent}}) \in \mathbb{R}^{|\mathcal{V}_{\tau}| \times d}
    \label{eq:state_mix}
\end{equation}

Concatenate with the broadcasted system-level information to form the enriched representation

\begin{equation}
    \mathcal{E}_{\text{mixing}} = \text{concat}\left( \mathcal{H}^{\text{env}}, \mathcal{H}_{\text{mixing}}^{V} \right) \in \mathbb{R}^{|\mathcal{V}_{\tau}| \times (d_{1} + d)}
\end{equation}

MLPs are then used to generate the weights and biases, as defined in Eq.~\ref{eq:weight} and Eq.~\ref{eq:bias}, respectively, with the MLPs applied row-wise to the representation $\mathcal{E}_{\text{mixing}}$.

\begin{align}
    \label{eq:weight}
    & \Big( w_v \Big)_{v \in \mathcal{V}_\tau} = \text{MLP}_{\text{weight}} (\mathcal{E}_{\text{mixing}}) \in \mathbb{R}^{|\mathcal{V}_{\tau}| \times 1} \\
    \label{eq:bias}
    & \Big( b_v \Big)_{v \in \mathcal{V}_\tau} = \text{MLP}_{\text{bias}} (\mathcal{E}_{\text{mixing}}) \in \mathbb{R}^{|\mathcal{V}_{\tau}| \times 1}
\end{align}

Afterwards, the total value function is computed as shown in Eq.~\ref{eq:total_value}. To satisfy the monotonicity constraint, i.e., $\frac{\partial f}{\partial Q_{v}} \ge 0$ for $\forall v \in \mathcal{V}_{\tau}$, the absolute values of the generated weights are used, ensuring that increases in individual agent values do not decrease the global value.

\begin{equation}
    Q_{\text{tot}} (s_{\tau}, \boldsymbol{a}_{\tau}) = \sum_{v \in \mathcal{V}_{\tau}} \left( |w_{v}| \times Q_{v} + b_{v} \right)
    \label{eq:total_value}
\end{equation}

It is clear that, due to the use of MHA layers, a variable number of agents can be processed, and an embedding for each agent can be obtained that captures semantic relationships with other agents. Moreover, since the MLP layers operate in a row-wise manner, they independently process each agent's embedding to produce the corresponding action values, weights, or biases, regardless of the number of agents. As a result, both the agent and the mixing network are capable of handling a variable agent set $\mathcal{V}_{\tau}$.

\subsubsection{Action Homogenization Mitigation}

In practice, for implementation efficiency, policies share parameters, which leads to action homogenization. To alleviate this issue, a lightweight yet effective mechanism is incorporated into \AVD, targeting the sources of highly similar observations and representations among agents.

\paragraph{Dispersed agent initialization} 
Action homogenization is particularly likely when multiple agents start from or operate in highly similar local conditions. To reduce this effect at the beginning of each episode, agents are initialized in a spatially dispersed manner. Specifically, at the initial time step, we cluster task-relevant locations into $|\mathcal{V}_1|$ groups, and initialize each agent at the center of one cluster. 
In the bike redistribution task considered in this work, stations are clustered based on spatial proximity using a $k$-nearest-neighbor-based clustering procedure, and each vehicle is initialized at the centroid of a distinct cluster. 
This initialization strategy encourages agents to experience diverse local conditions from the outset, reducing the likelihood of identical observations and subsequent homogenized actions.

\paragraph{Stochastic perturbation of agent embeddings} 
Even with dispersed initialization, agents may still encounter similar states during execution, especially when responding to concentrated demand patterns. To further promote action diversity without breaking policy sharing, small stochastic perturbations are injected into the agent embeddings produced by the agent network.
Concretely, after obtaining the contextualized agent embeddings $\mathcal{H}_{\text{agent}}^{V}$, Gaussian noise is added as below:
\begin{equation}
    \mathcal{H}_{\text{agent}}^{V}
    \leftarrow \mathcal{H}_{\text{agent}}^{V} + \boldsymbol{e},
    \quad \boldsymbol{e} \sim \mathcal{N}(0, \mathbf{I})
\end{equation}
after which, the perturbed embeddings are used in the subsequent processing procedure. This perturbation introduces mild stochasticity into the decision process, encouraging agents with similar embeddings to explore different actions, while preserving the underlying shared policy structure.

Together, the above two components constitute the homogenization mitigation mechanism, improving behavioral diversity among concurrently acting agents and mitigating action homogenization while preserving the efficiency benefits of parameter sharing.

\subsection{CTDE-based Training-Execution Strategy for Semi-MARL}

In many real-world MAS, actions may require different amounts of time to complete, leading to asynchronous agent behaviors. We develop a CTDE-based training–execution strategy in which temporally extended actions are represented as step-level transitions in the training data.

At each time step $\tau$, the current state $s_{\tau}$ is observed.  
For each active agent $v \in \mathcal{V}_\tau$, if its previous action has been completed, a new action is selected according to an $\epsilon$-greedy policy:

\begin{equation}
    a_{\tau}^{v} =
    \begin{cases}
        \displaystyle \operatorname*{argmax}_{a} Q_v(s_{\tau}, a), & \text{with probability } 1 - \epsilon,\\[2mm]
        \text{random action } a \in A, & \text{with probability } \epsilon.
    \end{cases}
    \label{eq:action_eps}
\end{equation}
If the previous action is still ongoing, the agent continues executing it, i.e., $a_{\tau}^{v} = a_{\tau -1}^{v}$.  
Once all individual actions are determined, the joint action is formed as $\boldsymbol{a}_{\tau} = (a_{\tau}^{v})_{v \in \mathcal{V}_\tau}$. Executing $\boldsymbol{a}_{\tau}$ in the environment yields the next state $s_{\tau + 1}$ and a global reward $r_{\tau}$. The resulting step-level transition $(s_{\tau}, \boldsymbol{a}_{\tau}, s_{\tau+1}, r_{\tau})$ is stored in the replay buffer. By repeating this process across all time steps, we obtain a trajectory of transitions $\left\{ (s_{\tau}, \boldsymbol{a}_{\tau}, s_{\tau + 1}, r_{\tau}) \right\}_{\tau = 1}^{H-1}$, capturing both asynchronous action execution and the corresponding rewards, which is then used to train \AVD.

\paragraph{Centralized training}
During training, the agent network and the mixing network are optimized based on the Bellman equation~\cite{DQN, atari_dqn}, as shown in Eq.~\ref{eq:bellman}, where $\Bar{Q}_{\text{tot}}$ denotes the target network that is periodically copied from $Q_{\text{tot}}$ to stabilize training~\cite{DDQN}.

\begin{small}
\begin{equation}
    \operatorname*{min} \sum_{(s_{\tau}, \boldsymbol{a}_{\tau}, s_{\tau +1}, r_{\tau})} \left( Q_{\text{tot}} (s_{\tau}, \boldsymbol{a}_{\tau}) - (r_{\tau} + \gamma \max_{\boldsymbol{a}_{\tau+1}} \bar{Q}_{\text{tot}} (s_{\tau +1}, \boldsymbol{a}_{\tau +1})) \right)^{2}
\label{eq:bellman}
\end{equation}
\end{small}

\paragraph{Decentralized execution}
During execution, each agent selects actions independently at each time step using its corresponding agent network $\left\{ Q_{v}|v \in \mathcal{V}_{\tau} \right\}$, as shown in Eq.~\ref{eq:action_eps}, or continues executing its previous action. Although actions are chosen in a decentralized manner, the monotonicity constraint imposed by the mixing network in Eq.~\ref{eq:qmix} ensures that independent action selection can still achieve cooperative optimality.

\begin{algorithm}[t]
    \SetKwInOut{Input}{Input}
    \SetKwInOut{Initial}{Initial}
    \SetKwInOut{Output}{Output}
    
    \Input{Number of steps per episode $H$; number of training episodes $N$; active dispatch vehicles $V$; exploration rate $\epsilon \in [0, 1]$; replay buffer $\mathcal{B} \leftarrow \emptyset$; number of training iterations $J$, soft target update rate $\alpha \in [0, 1]$}
    \vspace{0.2em}
    
    \Initial{Agent network $Q_{v}$ for $\forall v \in \mathcal{V}$ and mixing network $f$ to form the joint action-value network $Q_{\text{tot}}$; the target network $\Bar{Q}_{\text{tot}} \leftarrow Q_{\text{tot}}$}
    \vspace{0.2em}
    
	\Output{Learned agent network $Q_{v}$ for $\forall v \in \mathcal{V}$}
    \BlankLine 

    \For{\textup{episode} $n = 1$ to $N$}{
        Reset the environment \\
        \For{\textup{time step} $\tau = 1$ to $H$}{
            Get the current state $s_{\tau}$ \\
            \textcolor{gray}{\textit{// Semi-MARL training-execution strategy}} \\
            \For{\textup{agent} $v \in \mathcal{V}_\tau$}{
                \If{\textup{agent} $v$ \textup{is idle}}{
                    Generate action $a_{\tau}^{v}$ for $v$ by Eq.~\ref{eq:action_eps} \\
                }
                \Else{
                    Keep previous action $a_{\tau}^{v} = a_{\tau -1}^{v}$ for agent $v$ \\
                }
            }
            Execute the joint action $\boldsymbol{a}_{\tau} = (a_{\tau}^{1}; \cdots; a_{\tau}^{|\mathcal{V}_\tau|})$ \\
            Transition to state $s_{\tau + 1}$ and get a global reward $r_{\tau}$ \\
            Restore the sample $\mathcal{B} \leftarrow \mathcal{B} \cup \left\{ (s_{\tau}, \boldsymbol{a}_{\tau}, s_{\tau + 1}, r_{\tau}) \right\}$ \\
        }
        \textcolor{gray}{\textit{// \AVD with homogenization mitigation mechanism}} \\
        \For{$j = 1$ to $J$}{
        Sample a binimatch from $\mathcal{B}$ \\
        Update $Q_{\text{tot}}$ using the minibatch by Eq.~\ref{eq:bellman} \\
        }
        Update target network $\Bar{Q}_{\text{tot}} \leftarrow \alpha \times Q_{\text{tot}} + (1- \alpha) \times \Bar{Q}_{\text{tot}}$ \\
    }
    \Return{\textup{Agent network} $Q_{v}$ \textup{for} $\forall v \in \mathcal{V}$}
    \caption{\AVD Algorithm}
    \label{alg:avd}
\end{algorithm}

Alg.~\ref{alg:avd} summarizes the training–execution procedure for the proposed \AVD algorithm, illustrating how policies are learned from step-level transitions while preserving ongoing actions during execution. \textbf{Algorithmic parameters are provided in Tab.~\ref{tab:hyperparameters} in the Appendix.}

\begin{figure*}[t]
  \centering
  \begin{subfigure}[t]{0.33\textwidth}
    \centering
    \includegraphics[width=\linewidth, height=30mm, trim=0.3cm 0cm 0cm 0cm, clip]{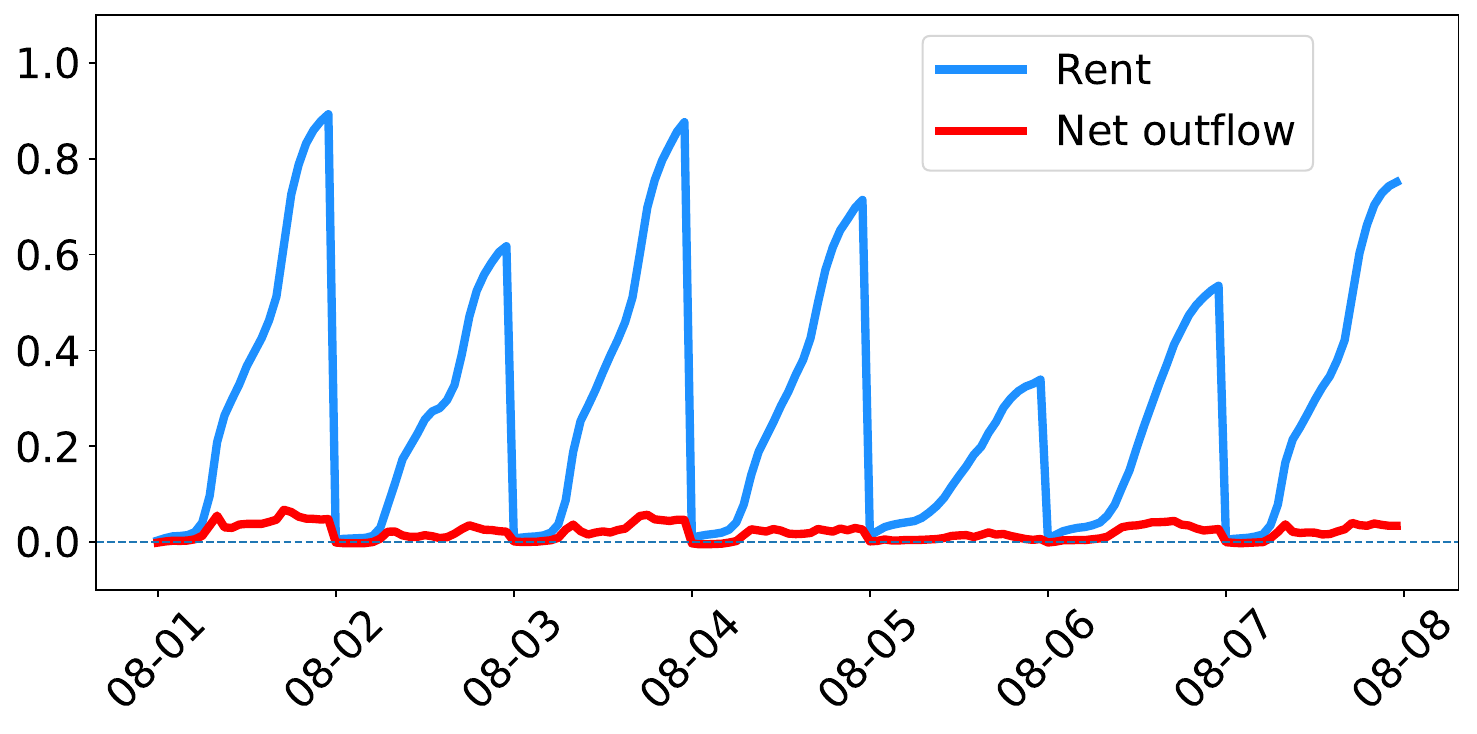}
    \caption{\texttt{Region 1 in London}}
    \label{fig:region_1}
  \end{subfigure}
  \hfill
  \begin{subfigure}[t]{0.33\textwidth}
    \centering
    \includegraphics[width=\linewidth, height=30mm, trim=0.3cm 0cm 0cm 0cm, clip]{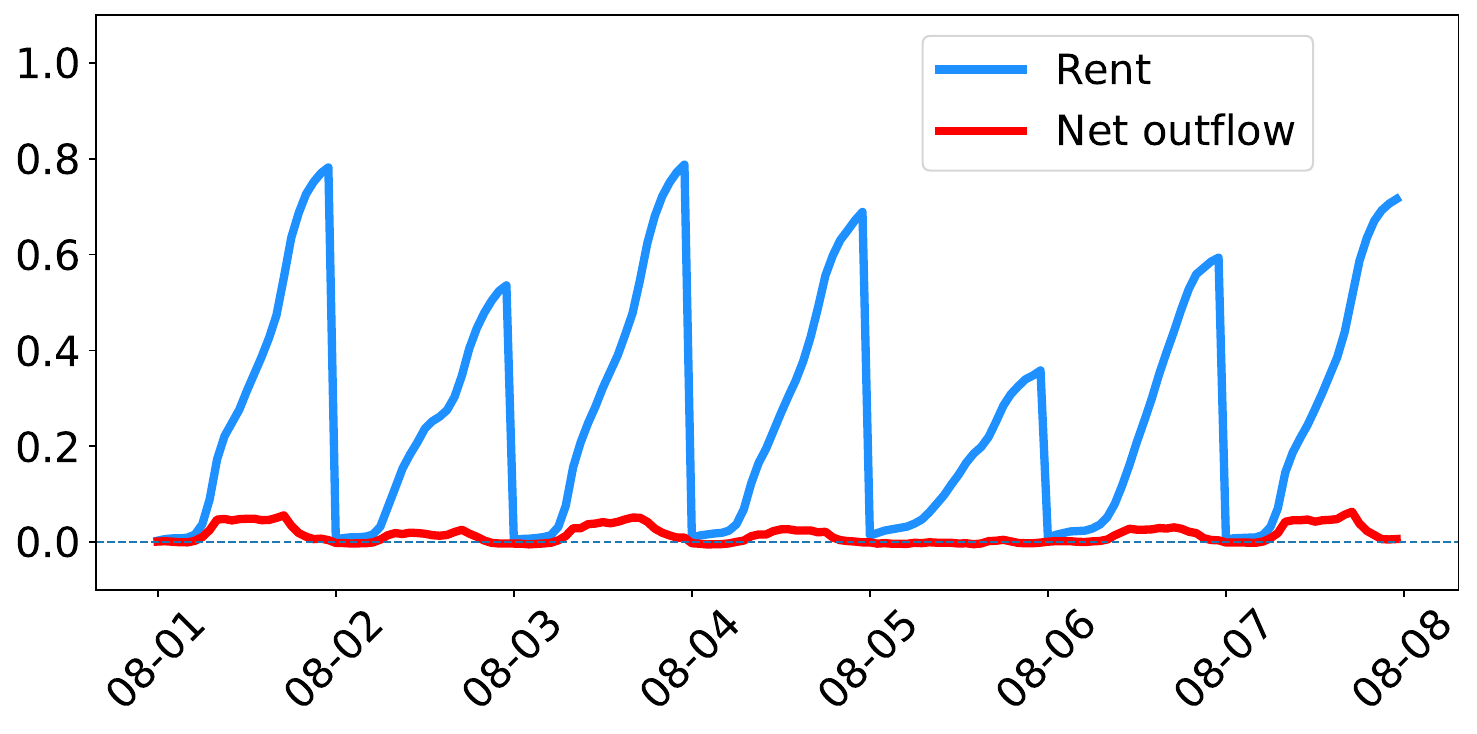}
    \caption{\texttt{Region 3 in London}}
    \label{fig:region_3}
  \end{subfigure}
  \hfill
  \begin{subfigure}[t]{0.33\textwidth}
    \centering
    \includegraphics[width=\linewidth, height=30mm, trim=0.3cm 0cm 0cm 0cm, clip]{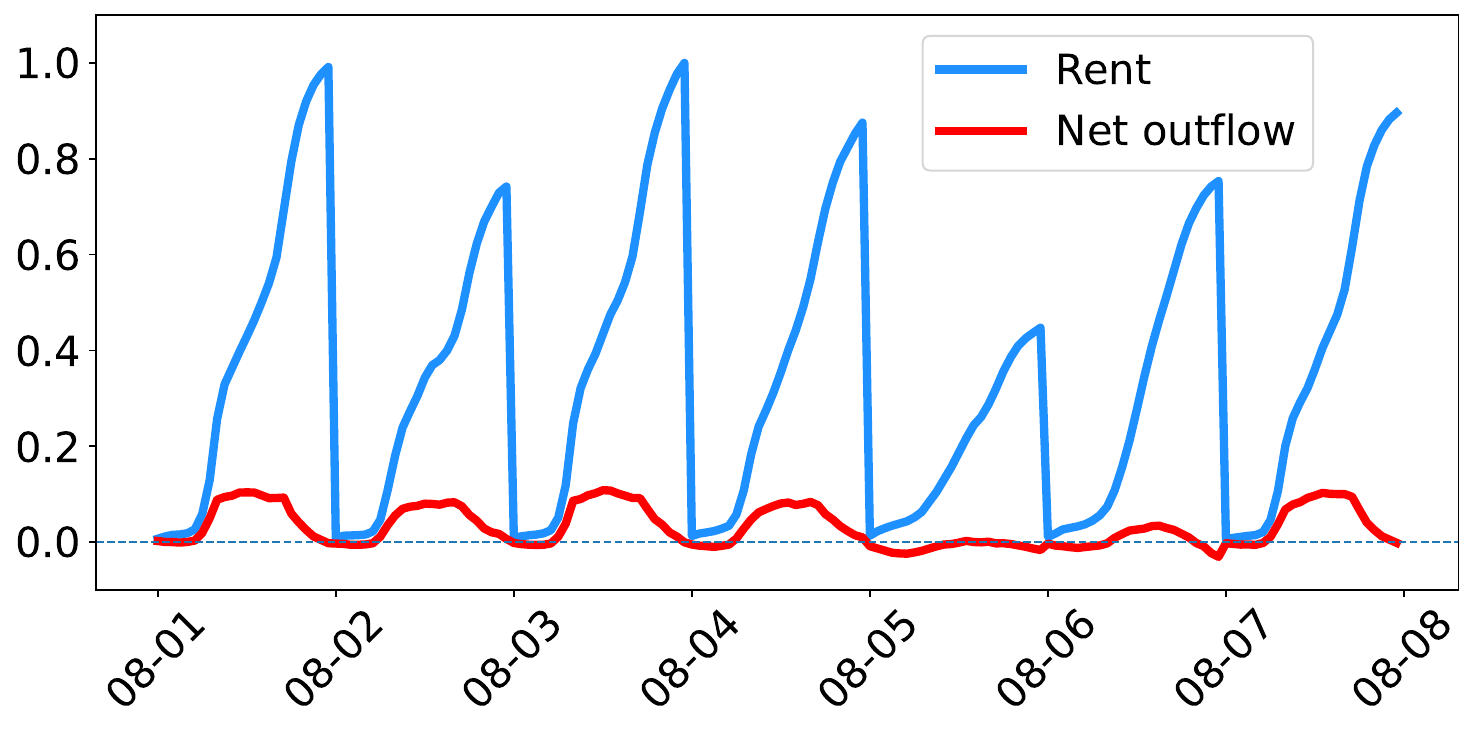}
    \caption{\texttt{Region 5 in London}}
    \label{fig:region_5}
  \end{subfigure}
  \hfill
  \caption{Daily cumulative net outflows remain minimal relative to cumulative trip volumes. Both bike rentals and net outflows are normalized by the maximum rental count across the regions to highlight the scale difference between the two measures.}
  \label{fig:flow}
\end{figure*}

\section{Experiment}

In this section, we evaluate the performance of \AVD using datasets from two real-world bike-sharing systems, i.e., the \textit{Santander Cycles} \footnote{\href{https://www.kaggle.com/code/raitest/london-bike-rental-analysis-august-2023}{Bike usage data} and \href{https://santandercycles.tfl.gov.uk/map}{station map} in London} in London and the \textit{Capital Bikeshare} \footnote{\href{https://capitalbikeshare.com/system-data}{Bike usage data} and \href{https://capitalbikeshare.com/}{station map} in Washington D.C.} in Washington D.C. Characteristics of both datasets are summarized in Tab.~\ref{tab:USR}.

\begin{table}[htbp]
\caption{Dataset statistics. Avg. duration refers to the average time between a bike being rented and returned.}
\centering
\resizebox{\columnwidth}{!}{
\begin{tabular}{ccccc}
\toprule
BSS & Time span & \# Regions & \# Stations & Avg. duration / min \\
\midrule
Santander Cycles & Aug. 2023 & 11 & 792 & 25.27\\
Capital Bikeshare & Jan. 2025 & 8 & 393 & 11.28\\ \bottomrule
\end{tabular}
}
\label{tab:USR}
\end{table}

\subsection{Simulator}
Since RL relies on trial-and-error, an environment simulator $\mathcal{S}$ is essential for effective training and evaluation. We develop a simulator that can be easily instantiated with historical data from any bike-sharing system, enabling seamless simulation of the corresponding operations. The simulator $\mathcal{S}$ comprises two main components.


\paragraph{Bike usage simulation} At each time step $\tau$, the simulator $\mathcal{S}$ processes bike usage records sequentially. For each record, if a bike is requested from a station, $\mathcal{S}$ checks whether any bikes are available there. If so, it decreases the number of bikes at that station by one, increments the cumulative count of successful rentals by one, and starts tracking the rented bike until it is returned. Otherwise, the rental request is considered lost and is excluded from further tracking. After processing rental requests, $\mathcal{S}$ checks whether any previously rented bikes are due to be returned to stations. If so, it updates the number of bikes at the corresponding stations accordingly and stops tracking these records.

\paragraph{Redistribution simulation} After simulating bike usage at $\tau$, the simulator $\mathcal{S}$ proceeds to simulate redistribution actions. For each vehicle that arrives at its destination, $\mathcal{S}$ verifies the feasibility of the assigned action and adjusts it if necessary. For example, if a vehicle is instructed to load 5 bikes but only 4 are available at the station, it loads the remaining 4 bikes instead. Afterward, the algorithm assigns a new redistribution action to the vehicle. The simulator estimates the action duration based on the Euclidean distance between the origin and destination stations and the vehicle’s speed, and continues tracking the vehicle until the task is completed.

\subsection{Experimental Settings}

\textbf{Baselines}. For comparison, three heuristic and three representative RL-based methods are adopted as baselines.
\begin{itemize}[leftmargin=1.5em]
\item \textbf{No Action}. No bike redistribution is performed; thus, the system relies solely on natural bike movements by users.
\item \textbf{Cooperative \texttt{Greedy}}. Stations are ranked by net rental intensity, with bikes loaded at the stations with the lowest intensity and unloaded at those with the highest intensity. By accounting for the actions of other agents, the algorithm exhibits a degree of cooperative behavior.

\item \textbf{\texttt{OPT}}~\cite{liu2016bike}. A heuristic approach based on an optimization model, where stations are partitioned into $|\mathcal{V}_1|$ clusters, and each agent is responsible for redistributing bikes within its assigned cluster.

\item \textbf{IDQN}~\cite{DQN}. Each agent independently trains its deep Q-network without any coordination or communication with other agents.

\item \textbf{VDN+}~\cite{sunehag2017value}. Constructs a total value function as the sum of individual agent value functions. This simple additive decomposition enables cooperative decision-making without relying on a mixing network. Moreover, its agent network is designed the same way as those in \AVD, making it more capable.

\item \textbf{QMIX}~\cite{rashid2020monotonic}. Constructs the total value function using a non-linear mixing network that combines individual agent Q-values, with the network constrained to be monotonic. However, it does not support adaptation to a varying number of agents.
\end{itemize}
\textbf{For a fair comparison, all RL-based baselines are augmented with our proposed homogenization-mitigation mechanism}. Its effectiveness is further validated through a dedicated ablation study presented later.

\renewcommand{\arraystretch}{1.1}
\begin{table}[H]
\caption{Algorithm comparison}
\label{tab:algorithm_comp}
\begin{center}
\resizebox{\linewidth}{!}{
\begin{tabular}{c|
                >{\centering\arraybackslash}m{1.5cm}
                >{\centering\arraybackslash}m{1.0cm}
                >{\centering\arraybackslash}m{0.8cm}
                >{\centering\arraybackslash}m{0.8cm}
                >{\centering\arraybackslash}m{0.8cm}
                >{\centering\arraybackslash}m{0.8cm}
                >{\centering\arraybackslash}m{0.8cm}
}
    \toprule
    Algorithm & No Action & \texttt{Greedy} & \texttt{OPT} & IDQN & VDN+ & QMIX & \cellcolor{blue!10} \AVD \\
    \midrule
    Adaptive & - & \cmark & \cmark & \cmark & \cmark & \xmark & \cellcolor{blue!10} \cmark \\
    \bottomrule
\end{tabular}
}
\end{center}
\end{table}
\renewcommand{\arraystretch}{1.0}

In Tab.~\ref{tab:algorithm_comp}, we summarize the capability of each algorithm to handle a varying number of agents. 
As shown, \texttt{Greedy}, \texttt{OPT}, IDQN, VDN+, as well as our proposed \AVD, can be applied to tasks with a dynamic agent population, whereas QMIX cannot.

\textbf{Evaluation metric}. In our experiments, we use the cumulative number of bike rentals per episode as the evaluation metric, which reflects how much rental demand is successfully served.

\textbf{Simulator settings}. The capacity of each vehicle is set to 5. At each time step, when generating actions, each agent considers the 15 nearest stations as candidate destinations. Accordingly, the action of agent $v$ at time step $\tau$ is defined as $a_{\tau}^{v} = (g_{\tau}^{v}, m_{\tau}^{v})$, where $g_{\tau}^{v}$ is the selected destination among the 15 nearest stations, and $m_{\tau}^{v} \in \left\{ -5, -4, -3, -2, -1, 0, 1, 2, 3, 4, 5 \right\}$. This yields an action space of size $15 \times 11 = 165$ for each agent. Once an action is assigned to an agent, its execution time is simulated based on the travel distance and a nominal vehicle speed of 12 km/h. Episodes are defined over two demand periods, from 7:00 AM to 1:00 PM for the morning period and from 1:00 PM to 8:00 PM for the afternoon period, to account for the noon break and shift changes. Each time step corresponds to 10 minutes.

\textbf{Redistribution within Administrative Regions}.  
Analysis of historical data indicates that individual administrative regions largely exhibit \textbf{self-balancing} bike flows, with net inflows and outflows remaining minimal relative to total trip volumes. Beyond this inherent self-balancing property, bike redistribution vehicles are typically managed on a per-region basis for operational convenience. Consequently, we partition each city according to administrative boundaries\footnote{\href{https://data.london.gov.uk/dataset/statistical-gis-boundary-files-london/}{GIS boundary files}} and assign multiple vehicles to operate within each region. The redistribution algorithm coordinates these vehicles only within their assigned region, without considering vehicles in other regions. Fig.~\ref{fig:London} and Fig.~\ref{fig:Washington} in the Appendix illustrate these regions, while Fig.~\ref{fig:flow} compares cumulative bike rentals and net outflows in three representative regions, further highlighting their near self-balancing behavior.

\subsection{Overall Performance}

\renewcommand{\arraystretch}{1.2}
\begin{table*}[t]
\caption{Experimental results on the London dataset. At the beginning of each episode, initialize the inventory levels of each station using a predefined ratio of 20\% and 30\%. \texttt{Overall} indicates the total return aggregated across all regions.}
\label{tab:experiment_result}
\centering
\resizebox{\linewidth}{!}{
    \begin{tabular}{c|c|
                        >{\centering\arraybackslash}m{1.0cm}|
                        >{\centering\arraybackslash}m{1.0cm}|
                        >{\centering\arraybackslash}m{1.0cm}|
                        >{\centering\arraybackslash}m{1.0cm}|
                        >{\centering\arraybackslash}m{1.0cm}|
                        >{\centering\arraybackslash}m{1.0cm}|
                        >{\centering\arraybackslash}m{1.0cm}|
                        >{\centering\arraybackslash}m{1.0cm}|
                        >{\centering\arraybackslash}m{1.0cm}|
                        >{\centering\arraybackslash}m{1.0cm}|
                        >{\centering\arraybackslash}m{1.0cm}|
                        >{\centering\arraybackslash}m{1.0cm}
    }

        \toprule
        
        \multicolumn{2}{c|}{\diagbox{Algorithm}{\texttt{Env.}}} 
        & \multicolumn{2}{c|}{\texttt{Region 1}}  
        & \multicolumn{2}{c|}{\texttt{Region 2}} 
        & \multicolumn{2}{c|}{\texttt{Region 3}} 
        & \multicolumn{2}{c|}{\texttt{Region 4}}
        & \multicolumn{2}{c|}{\texttt{Region 5}} 
        & \multicolumn{2}{c}{\texttt{Overall}} \\
        \midrule
        

        \multicolumn{2}{c|}{Initial inventory ratio} 
        & 20\% & 30\% & 20\% & 30\% & 20\% & 30\% & 20\% & 30\% & 20\% & 30\% & 20\% & 30\% \\
        \midrule

        \multicolumn{2}{c|}{No Action} 
        & 1287 & 1512 & 1553 & 1725 & 1340 & 1514 & 1054 & 1294 & 1531 & 1813 & 6765 & 7858 \\
        \midrule
        
        \multirow{6}{*}{\shortstack{$|\mathcal{V}_\tau| \equiv 2$}} & 
        \texttt{Greedy}
        & 1307 & 1540 & 1594 & 1753 & 1387 & 1546 & 1087 & 1332 & 1561 & 1835 & 6936 & 8006 \\
        & \texttt{OPT}
        & 1330 & 1563 & 1592 & \underline{1771} & 1384 & 1552 & 1099 & 1343 & 1577 & 1850 & 6982 & 8079 \\
        & IDQN
        & 1333 & 1563 & 1608 & 1747 & 1394 & 1544 & 1112 & 1343 & 1576 & 1842 & 7023 & 8039 \\
        & VDN+
        & 1345 & 1571 & \underline{1618} & 1770 & 1411 & 1574 & 1112 & \underline{1361} & \underline{1614} & 1869 & 7100 & 8145 \\
        & QMIX
        & \underline{1356} & \underline{1590} & 1588 & 1750 & \underline{1429} & \underline{1582} & \underline{1129} & 1352 & 1612 & \underline{1890} & \underline{7114} & \underline{8164} \\

        & \cellcolor{blue!10} AVD
        & \cellcolor{blue!10} \textbf{1366} & \cellcolor{blue!10} \textbf{1595} & \cellcolor{blue!10} \textbf{1641} & \cellcolor{blue!10} \textbf{1784} & \cellcolor{blue!10} \textbf{1448} & \cellcolor{blue!10} \textbf{1587} & \cellcolor{blue!10} \textbf{1140} & \cellcolor{blue!10} \textbf{1370} & \cellcolor{blue!10} \textbf{1629} & \cellcolor{blue!10} \textbf{1905} & \cellcolor{blue!10} \textbf{7224} & \cellcolor{blue!10} \textbf{8241} \\

        \midrule

        \multirow{6}{*}{\shortstack{$|\mathcal{V}_\tau| \equiv 4$}} & 
        \texttt{Greedy}
        & 1346 & 1579 & 1621 & 1791 & 1403 & 1569 & 1106 & 1359 & 1586 & 1871 & 7062 & 8169 \\
        & \texttt{OPT}
        & 1349 & 1587 & \underline{1635} & \underline{1810} & \underline{1445} & \textbf{1622} & 1117 & 1370 & 1626 & 1877 & 7172 & 8266 \\
        & IDQN
        & 1339 & 1569 & 1610 & 1773 & 1399 & 1526 & 1107 & 1333 & 1582 & 1847 & 7037 & 8048 \\
        & VDN+
        & 1352 & 1596 & 1624 & 1777 & \underline{1445} & 1575 & \underline{1142} & \underline{1402} & \underline{1656} & \underline{1927} & \underline{7219} & \underline{8277} \\
        & QMIX
        & \underline{1369} & \underline{1614} & 1633 & 1783 & 1415 & 1579 & 1139 & 1358 & 1617 & 1900 & 7173 & 8234 \\
        
        & \cellcolor{blue!10} AVD
        & \cellcolor{blue!10} \textbf{1382} & \cellcolor{blue!10} \textbf{1639} & \cellcolor{blue!10} \textbf{1661} & \cellcolor{blue!10} \textbf{1818} & \cellcolor{blue!10} \textbf{1463} & \cellcolor{blue!10} \underline{1610} & \cellcolor{blue!10} \textbf{1167} & \cellcolor{blue!10} \textbf{1420} & \cellcolor{blue!10} \textbf{1662} & \cellcolor{blue!10} \textbf{1958} & \cellcolor{blue!10} \textbf{7335} & \cellcolor{blue!10} \textbf{8445} \\

        \midrule
        \multicolumn{14}{c}{\#Agents varies} \\

        \midrule
        
        \multirow{3}{*}{\shortstack{$|\mathcal{V}_\tau|=4 \to 2 $}} & 
        \texttt{OPT}
        & 1332 & 1574 & 1616 & \underline{1792} & \underline{1428} & \textbf{1614} & 1110 & 1364 & 1609 & 1864 & 7095 & 8208 \\
        & VDN+
        & \underline{1360} & \underline{1575} & \underline{1619} & 1771 & 1422 & 1566 & \underline{1145} & \underline{1394} & \underline{1617} & \underline{1918} & \underline{7163} & \underline{8223} \\
        & \cellcolor{blue!10} AVD
        & \cellcolor{blue!10} \textbf{1385} & \cellcolor{blue!10} \textbf{1634} & \cellcolor{blue!10} \textbf{1659} & \cellcolor{blue!10} \textbf{1817} & \cellcolor{blue!10} \textbf{1455} & \cellcolor{blue!10} \underline{1605} & \cellcolor{blue!10} \textbf{1163} & \cellcolor{blue!10} \textbf{1409} & \cellcolor{blue!10} \textbf{1653} & \cellcolor{blue!10} \textbf{1942} & \cellcolor{blue!10} \textbf{7315} & \cellcolor{blue!10} \textbf{8407} \\
        \bottomrule
    \end{tabular}
}
\end{table*}
\renewcommand{\arraystretch}{1.0}

For brevity, we \textbf{denote $\equiv$ as \textit{identically equal}} in the following. As shown in Table~\ref{tab:experiment_result}, when the number of agents per region is fixed across time steps, i.e., $|\mathcal{V}_\tau| \equiv 2$ or $|\mathcal{V}_\tau| \equiv 4$ for $\tau = 1, 2, \ldots, H$, \AVD significantly outperforms all baselines. Notably, VDN+ and QMIX do not consistently achieve higher cumulative returns as the number of agents increases, whereas \AVD maintains stable performance gains. IDQN performs the worst among all methods and even underperforms the heuristic baseline \texttt{Greedy}, which incorporates limited inter-agent coordination, while IDQN operates independently across agents. Moreover, IDQN’s performance further deteriorates as the number of agents increases, indicating its limited ability to support multi-agent coordination. Interestingly, \texttt{OPT} occasionally surpasses \AVD when the number of active agents is fixed at $|\mathcal{V}_\tau| \equiv 4$. Although infrequent, this result is noteworthy and likely arises because \texttt{OPT} assigns agents to separate working areas, thereby reducing inter-agent conflicts at higher agent counts. Overall, these results suggest that coordination becomes increasingly challenging as the number of agents grows, and demonstrate that \AVD is more effective in enabling cooperation among agents.

In addition, we evaluate a dynamic-agent setting in which the number of active agents varies over time, with $|\mathcal{V}_\tau|$ set to 4 during rush hours from 7:00 AM to 10:00 AM and 2 during off-peak periods from 10:00 AM to 1:00 PM. We compare \AVD against the strongest heuristic and RL-based baselines capable of handling dynamic agent populations, namely \texttt{OPT} and VDN+. The results show that \AVD continues to outperform these baselines, demonstrating its robustness under varying agent populations.

Notably, although the time-averaged number of agents in this setting is 3, the performance of \AVD does not simply fall between the fixed-agent cases of 2 and 4. Instead, it significantly surpasses the performance with 2 agents and is only slightly below that with 4 agents. This observation indicates that dynamically adjusting the number of deployed agents—allocating more during peak periods and fewer during off-peak periods—can substantially improve overall returns under a fixed deployment cost, highlighting the practical advantage of \AVD in real-world urban systems.

\subsection{Robustness Validation}
Additionally, we validate the robustness of \AVD on the Washington, D.C. dataset, where overall bike usage is substantially lower than in London. Experiments were conducted in January, a period characterized by particularly low demand, and the initial number of bikes at each station was set to a reduced level. This setup increases the precision required for effective redistribution: with both demand and supply scarce, even minor redistribution errors can lead to stockouts, resulting in the loss of an already limited number of potential rentals. Consequently, the system becomes highly sensitive to suboptimal coordination and timing, providing a stringent test of \AVD's stability and robustness. As shown in Table~\ref{tab:experiment_result_dc}, \AVD continues to outperform all baselines under these challenging conditions. Interestingly, \texttt{OPT} and QMIX, which previously outperformed \texttt{Greedy} and VDN+ on the London dataset, are now respectively surpassed by \texttt{Greedy} and VDN+, highlighting the sensitivity of baseline methods to variations in demand and initialization settings.

\subsection{Zero-shot Performance}

\begin{figure}[t]
    \centering
    \includegraphics[width=\linewidth]{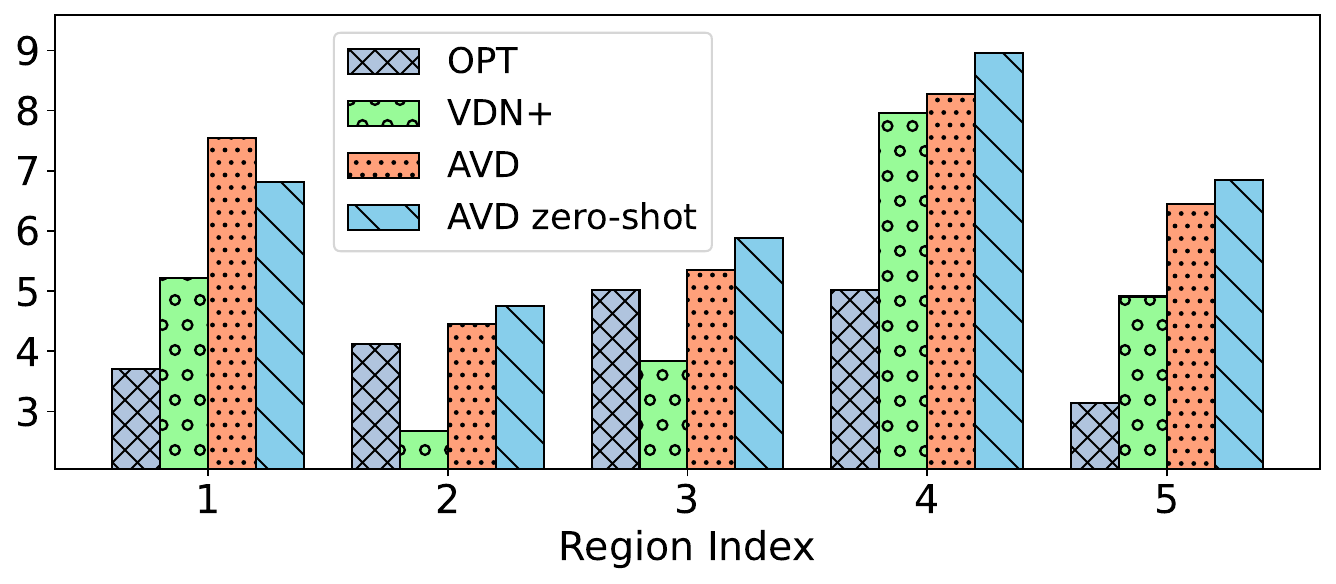}
    \caption{Zero-shot performance of \AVD trained with $|\mathcal{V}_\tau| \equiv 4$ and directly deployed with $|\mathcal{V}_\tau| \equiv 3$ without retraining, compared to \AVD and VDN+ trained with $|\mathcal{V}_\tau| \equiv 3$ and \texttt{OPT} with $|\mathcal{V}_\tau| \equiv 3$. In this setting, the initial inventory is set to 30\%, and the observed conclusions generalize to other configurations.}
    \label{fig:zero-shot}
\end{figure}

Since \AVD supports a varying number of agents, we evaluate its zero-shot performance on the London dataset. Specifically, the \AVD policy is trained with $|\mathcal{V}_\tau| \equiv 4$ agents and directly deployed in scenarios with only $|\mathcal{V}_\tau| \equiv 3$ agents at test time, without any retraining. For reference, we include comparisons with \AVD and VDN+ trained directly with $|\mathcal{V}_\tau| \equiv 3$ agents, as well as \texttt{OPT} with $|\mathcal{V}_\tau| \equiv 3$. For better visual clarity and comparability, Fig.~\ref{fig:zero-shot} presents the percentage improvement of each method relative to the \textit{No Action} baseline, while the exact numerical results are provided in Tab.~\ref{tab:zero-shot} in the Appendix.

As shown in Fig.~\ref{fig:zero-shot}, zero-shot \AVD achieves performance comparable to—and in some regions better than—the directly trained \AVD with three agents. Both variants consistently outperform the learning-based baseline VDN+ and the \texttt{OPT} method, demonstrating strong generalization capability. The occasional superiority of zero-shot \AVD stems from the richer interaction patterns learned during training with more agents, which provide a coordination prior that enhances robustness and reduces overfitting to a specific agent cardinality. When deployed with fewer agents, this prior enables more decisive redistribution behaviors, whereas models trained directly with fewer agents may overfit to limited coordination structures. These results suggest that \AVD learns coordination patterns that generalize across different agent cardinalities, rather than a policy tied to a fixed number of agents, which is advantageous for real-world settings with varying numbers of active agents.

\renewcommand{\arraystretch}{1.1}
\begin{table}[H]
\caption{Experimental results on the Washington D.C. dataset. At the beginning of each episode, station inventory levels are initialized to 5\%, 7\%, 10\%, and 15\% of capacity. \texttt{Avg.} denotes the average performance across different initialization settings.}
\label{tab:experiment_result_dc}
\centering
\resizebox{\linewidth}{!}{
    \begin{tabular}{
                        c|
                        c|
                        >{\centering\arraybackslash}m{0.8cm}|
                        >{\centering\arraybackslash}m{0.8cm}|
                        >{\centering\arraybackslash}m{0.8cm}|
                        >{\centering\arraybackslash}m{0.8cm}|
                        >{\centering\arraybackslash}m{0.8cm}
    }

        \toprule
        
        \multicolumn{2}{c|}{\diagbox{Algorithm}{\texttt{Env.}}} 
        & \multicolumn{5}{c}{\texttt{Region 2}} \\
        \midrule


        \multicolumn{2}{c|}{Initial inventory ratio} 
        & 5\% & 7\% & 10\% & 15\% & \texttt{Avg.} \\
        \midrule

        \multicolumn{2}{c|}{No Action} 
        & 220 & 244 & 314 & 364 & 289 \\
        \midrule
        
        \multirow{6}{*}{\shortstack{$|\mathcal{V}_\tau| \equiv 2$}} & 
        \texttt{Greedy}
        & 233 & 252 & 335 & \underline{394} & 304 \\
        & \texttt{OPT}
        & 227 & 254 & 332 & 381 & 299 \\
        & IDQN
        & 242 & 267 & 341 & 388 & 310 \\
        & VDN+
        & \underline{269} & \textbf{290} & \underline{344} & \underline{394} & \underline{324} \\
        & QMIX
        & 249 & 269 & 341 & 391 & 313 \\
        
        & \cellcolor{blue!10} AVD
        & \cellcolor{blue!10} \textbf{272}  & \cellcolor{blue!10} \textbf{290} & \cellcolor{blue!10} \textbf{357} & \cellcolor{blue!10} \textbf{408} & \cellcolor{blue!10} \textbf{332} \\

        \midrule

        \multirow{6}{*}{\shortstack{$|\mathcal{V}_\tau| \equiv 4$}} & 
        \texttt{Greedy}
        & 236 & 266 & 351 & \underline{398} & 313 \\
        & \texttt{OPT}
        & 226 & 250 & 335 & 386 & 302 \\
        & IDQN
        & 246 & 265 & 336 & 385 & 308 \\
        & VDN+
        & \underline{275} & \underline{297} & \underline{370} & 389 & \underline{333} \\
        & QMIX
        & 258 & 284 & 345 & 393 & 320 \\
        
        & \cellcolor{blue!10} AVD
        & \cellcolor{blue!10} \textbf{288} & \cellcolor{blue!10} \textbf{308} & \cellcolor{blue!10} \textbf{372} & \cellcolor{blue!10} \textbf{418} & \cellcolor{blue!10} \textbf{347} \\

        \midrule
        \multicolumn{7}{c}{\#Agents varies} \\

        \midrule

        \multirow{3}{*}{\shortstack{$|\mathcal{V}_\tau| = $ \\ $4 \to 2$}} & 
        \texttt{OPT} & 225 & 250 & 329 & 385 & 297 \\
        & VDN+
        & \underline{275} & \underline{293} & \underline{369} & \underline{394} & \underline{333} \\
        
        & \cellcolor{blue!10} AVD
        & \cellcolor{blue!10} \textbf{283} & \cellcolor{blue!10} \textbf{302} & \cellcolor{blue!10} \textbf{367} & \cellcolor{blue!10} \textbf{415} & \cellcolor{blue!10} \textbf{342} \\

        \bottomrule
    \end{tabular}
}
\end{table}
\renewcommand{\arraystretch}{1.0}

\subsection{Ablation Study}

\begin{figure}[t]
  \centering
  \begin{subfigure}[t]{0.23\textwidth}
    \centering
    \includegraphics[width=\linewidth, height=30mm, trim=0.3cm 0cm 0cm 0cm, clip]{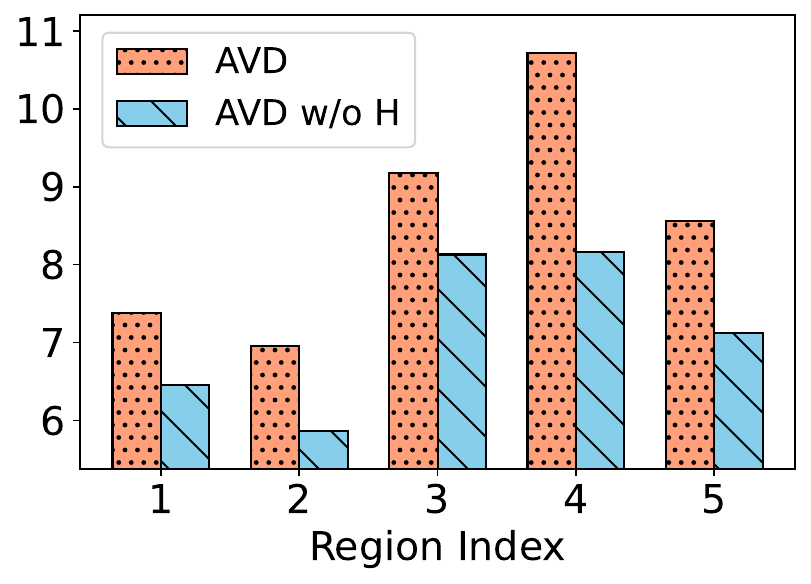}
    \caption{Initial inventory as 20\%}
    \label{fig:ab_20}
  \end{subfigure}
  \hfill
  \begin{subfigure}[t]{0.23\textwidth}
    \centering
    \includegraphics[width=\linewidth, height=30mm, trim=0.3cm 0cm 0cm 0cm, clip]{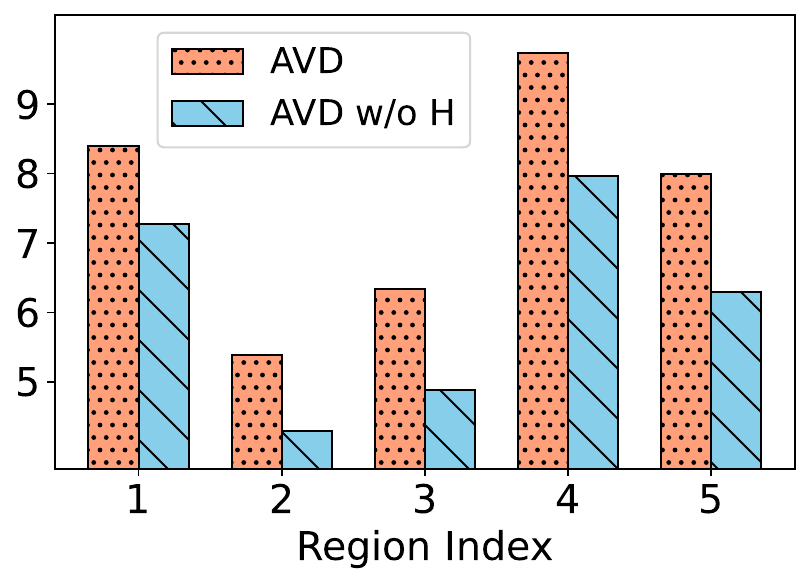}
    \caption{Initial Inventory as 30\%}
    \label{fig:ab_30}
  \end{subfigure}
  \hfill
  \caption{Ablation study of the lightweight mechanism for mitigating action homogenization. \AVD w/o H denotes the variant of \AVD without this mechanism. In this experiment, the number of agents is fixed at $|\mathcal{V}_\tau| \equiv 4$, and the observed conclusions generalize to other settings.}
  \label{fig:ablation}
\end{figure}

In this subsection, we conduct an ablation study on the London dataset to assess the effectiveness of the mechanism for mitigating action homogenization. For ease of visual comparison and consistent with the previous subsection, Fig.~\ref{fig:ablation} depicts the percentage improvement of each method relative to the \textit{No Action} baseline, while the numerical results are provided in Tab.~\ref{tab:ablation} in the Appendix.

As shown in Fig.~\ref{fig:ablation}, across all evaluated scenarios, incorporating the lightweight mechanism consistently improves performance compared to the variant without it, demonstrating its effectiveness in enhancing overall model behavior. Importantly, \textbf{these improvements are achieved with minimal computational overhead}, highlighting that introducing controlled diversity among agents can yield tangible gains without compromising efficiency, and thereby underscores the practical value of the homogenization-mitigation mechanism for urban MAS.

\section{Conclusion}

In this work, we study the challenges of coordinating urban multi-agent systems (MAS) in scenarios where the number of active agents varies over time, actions have heterogeneous durations, and shared policies can induce homogeneous behaviors. To address these challenges, we propose Adaptive Value Decomposition (\AVD), a cooperative MARL framework that adapts to a dynamically changing agent population and incorporates a lightweight mechanism to mitigate action homogenization. Furthermore, we design a training–execution strategy under the CTDE paradigm tailored for semi-MARL settings, enabling agents to make asynchronous decisions while maintaining coordinated behavior. Extensive experiments on real-world bike redistribution tasks in London and Washington, D.C., demonstrate that \AVD consistently outperforms representative baselines under both static and dynamic agent population scenarios. Finally, we release a high-fidelity bike-sharing simulator to provide a standardized benchmark for studying resource redistribution and multi-agent coordination in urban systems. Future work may explore other urban systems and larger-scale settings.

\bibliographystyle{ACM-Reference-Format}
\bibliography{sample-base}


\appendix

\renewcommand{\arraystretch}{1.2}
\begin{table*}[t]
\caption{Experimental results for the zero-shot evaluation.}
\label{tab:zero-shot}
\centering
\resizebox{0.68\linewidth}{!}{
    \begin{tabular}{c|c|
                        >{\centering\arraybackslash}m{1.3cm}|
                        >{\centering\arraybackslash}m{1.3cm}|
                        >{\centering\arraybackslash}m{1.3cm}|
                        >{\centering\arraybackslash}m{1.3cm}|
                        >{\centering\arraybackslash}m{1.3cm}|
                        >{\centering\arraybackslash}m{1.3cm}
    }

        \toprule
        
        \multicolumn{2}{c|}{\diagbox{Algorithm}{\texttt{Env.}}} 
        & \texttt{Region 1}
        & \texttt{Region 2}
        & \texttt{Region 3} 
        & \texttt{Region 4}
        & \texttt{Region 5}
        & \texttt{Overall} \\
        \midrule

        \multicolumn{2}{c|}{No Action} 
        & 1512 & 1725 & 1514 & 1294 & 1813 & 7858 \\
        \midrule
        
        \multirow{4}{*}{\shortstack{$|\mathcal{V}_\tau| \equiv 3$}} & 
        \texttt{OPT}
        & 1568 & 1796 & 1590 & 1359 & 1870 & 8183 \\
        & VDN+
        & 1591 & 1771 & 1572 & 1397 & 1902 & 8233 \\
        & \cellcolor{blue!10} \AVD
        & \cellcolor{blue!10} \textbf{1626} & \cellcolor{blue!10} \underline{1802} & \cellcolor{blue!10} \underline{1595} & \cellcolor{blue!10} \underline{1401} & \cellcolor{blue!10} \underline{1930} & \cellcolor{blue!10} \underline{8354} \\
        & \AVD zero-shot
        & 1615 & \textbf{1807} & \textbf{1603} & \textbf{1410} & \textbf{1937} & \textbf{8372} \\

        \bottomrule

    \end{tabular}
}
\end{table*}
\renewcommand{\arraystretch}{1.0}

\renewcommand{\arraystretch}{1.2}
\begin{table*}[t]
\caption{Experimental results of the ablation study.}
\label{tab:ablation}
\centering
\resizebox{\linewidth}{!}{
    \begin{tabular}{c|c|
                        >{\centering\arraybackslash}m{1.0cm}|
                        >{\centering\arraybackslash}m{1.0cm}|
                        >{\centering\arraybackslash}m{1.0cm}|
                        >{\centering\arraybackslash}m{1.0cm}|
                        >{\centering\arraybackslash}m{1.0cm}|
                        >{\centering\arraybackslash}m{1.0cm}|
                        >{\centering\arraybackslash}m{1.0cm}|
                        >{\centering\arraybackslash}m{1.0cm}|
                        >{\centering\arraybackslash}m{1.0cm}|
                        >{\centering\arraybackslash}m{1.0cm}|
                        >{\centering\arraybackslash}m{1.0cm}|
                        >{\centering\arraybackslash}m{1.0cm}
    }

        \toprule
        
        \multicolumn{2}{c|}{\diagbox{Algorithm}{\texttt{Env.}}} 
        & \multicolumn{2}{c|}{\texttt{Region 1}}  
        & \multicolumn{2}{c|}{\texttt{Region 2}} 
        & \multicolumn{2}{c|}{\texttt{Region 3}} 
        & \multicolumn{2}{c|}{\texttt{Region 4}}
        & \multicolumn{2}{c|}{\texttt{Region 5}} 
        & \multicolumn{2}{c}{\texttt{Overall}} \\
        \midrule

        \multicolumn{2}{c|}{Initial inventory ratio} 
        & 20\% & 30\% & 20\% & 30\% & 20\% & 30\% & 20\% & 30\% & 20\% & 30\% & 20\% & 30\% \\
        \midrule

        \multicolumn{2}{c|}{No Action} 
        & 1287 & 1512 & 1553 & 1725 & 1340 & 1514 & 1054 & 1294 & 1531 & 1813 & 6765 & 7858 \\
        \midrule
        
        \multirow{2}{*}{\shortstack{$|\mathcal{V}_\tau| \equiv 4$}} & 
        \texttt{\AVD w/o H}
        & 1370 & 1622 & 1644 & 1799 & 1449 & 1588 & 1140 & 1397 & 1640 & 1927 & 7243 & 8333 \\

        & \cellcolor{blue!10} \AVD
        & \cellcolor{blue!10} \textbf{1382} & \cellcolor{blue!10} \textbf{1639} & \cellcolor{blue!10} \textbf{1661} & \cellcolor{blue!10} \textbf{1818} & \cellcolor{blue!10} \textbf{1463} & \cellcolor{blue!10} \textbf{1610} & \cellcolor{blue!10} \textbf{1167} & \cellcolor{blue!10} \textbf{1420} & \cellcolor{blue!10} \textbf{1662} & \cellcolor{blue!10} \textbf{1958} & \cellcolor{blue!10} \textbf{7335} & \cellcolor{blue!10} \textbf{8445} \\

        \bottomrule

    \end{tabular}
}
\end{table*}
\renewcommand{\arraystretch}{1.0}

\vspace{50pt}
\section{Implementation Setting}

Algorithms are implemented in Python with PyTorch. Experiments were conducted on an Ubuntu 22.04 LTS system equipped with a 13th Gen Intel Core i9-13900KF CPU and an NVIDIA RTX 4090 GPU. Hyperparameters of \AVD are summarized in \autoref{tab:hyperparameters}.

\section{Experimental Results}

\begin{figure*}[t]
    \centering
    \includegraphics[width=0.8\linewidth, height=90mm, trim=0cm 0cm 0cm 0.74cm, clip]{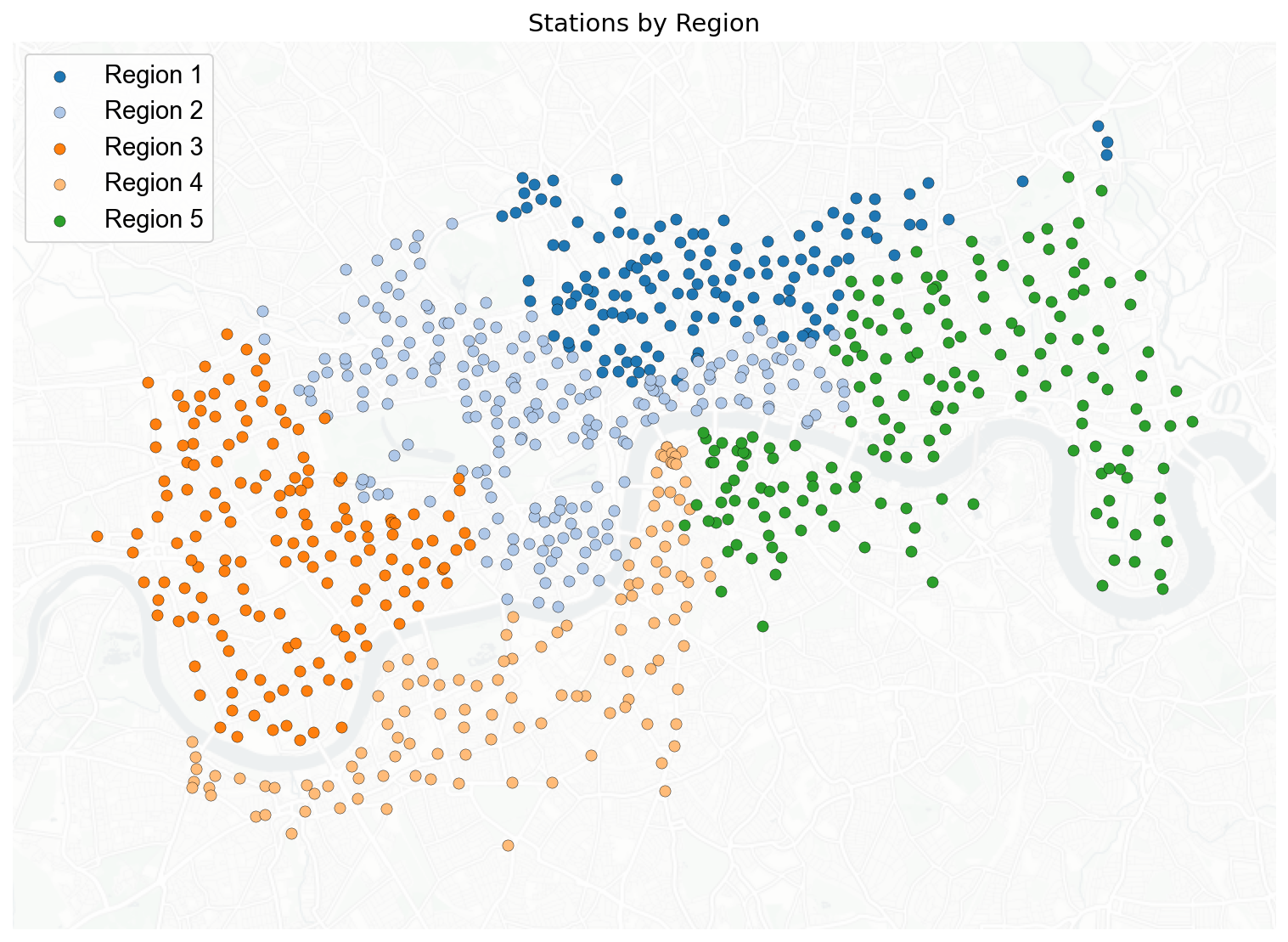}
    \caption{Map of London partitioned into 5 regions.}
    \label{fig:London}
\end{figure*}

\begin{figure*}[t]
    \centering
    \includegraphics[width=0.6\linewidth, height=95mm, trim=0cm 0cm 0cm 0.74cm, clip]{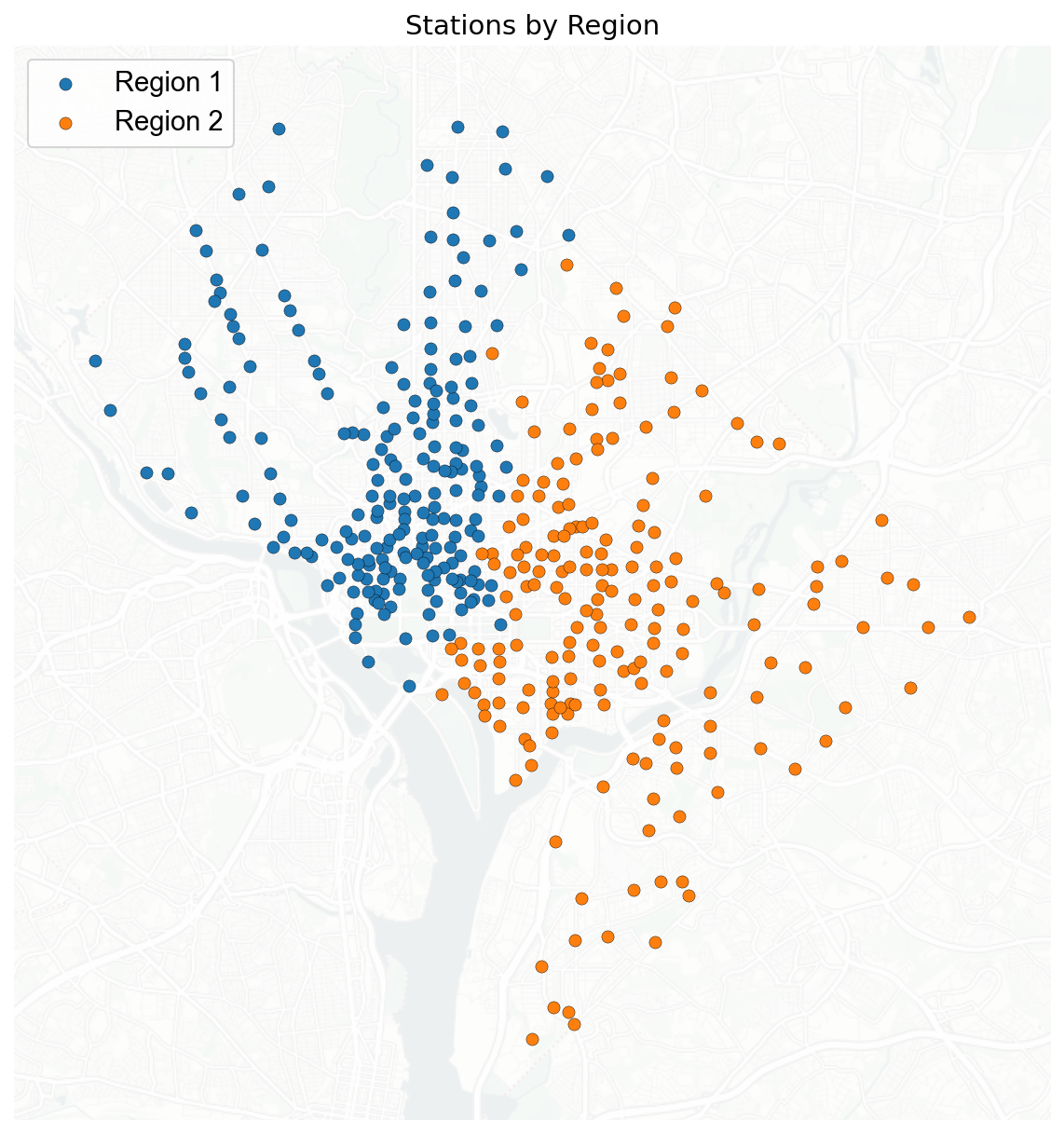}
    \caption{Map of Washington D.C. partitioned into 2 regions.}
    \label{fig:Washington}
\end{figure*}

In addition to the zero-shot evaluation and ablation study results shown in Fig.~\ref{fig:zero-shot} and Fig.~\ref{fig:ablation}, their corresponding numerical values are provided in Tab.~\ref{tab:zero-shot} and Tab.~\ref{tab:ablation}, respectively. Region maps of London and Washington D.C. are shown in Fig.~\ref{fig:London} and Fig.~\ref{fig:Washington}, respectively.

\renewcommand{\arraystretch}{1.3}
\begin{table}[H]
\caption{\AVD Hyperparameters}
\label{tab:hyperparameters}
\begin{center}
\resizebox{\linewidth}{!}{
\begin{tabular}{l|c|l|c}
    \toprule
    \multicolumn{2}{c|}{Policy training settings} & \multicolumn{2}{c}{Network architecture} \\
    \midrule
    Discounting factor $\gamma$ & $0.99$ & MHA output dimension $d$ & 32 \\
    \hline
    $\epsilon-$greedy rate & 0.1 & \# Heads & 1 \\
    \hline
    Number of training $J$  & $10$ & \# MHA layers & 1 \\
    \hline
    Learning rate & $5e-4$ & Dropout & 0.3 \\
    \hline
    Max gradient norm & $0.5$ & \# MLP layers & 2 \\
    \hline
    Soft target update rate $\alpha$  & $0.005$ & MLP hidden dimension & 64 \\
    \bottomrule
\end{tabular}
}
\end{center}
\end{table}
\renewcommand{\arraystretch}{1.0}

\end{document}